\documentclass[aps,prl,twocolumn,showpacs,superscriptaddress,groupedaddress]{revtex4}  
\usepackage{graphicx}  
\usepackage{dcolumn}   
\usepackage{bm}        
\usepackage{amssymb}   

\begin{document}


\title{Electron kinetics inferred from observations of microwave bursts during edge localised modes in the Mega-Amp Spherical Tokamak}
\author{S.J.~Freethy} \affiliation{CCFE, Culham Science Centre, Abingdon, OX14 3DB, UK}
\author{K.G.~McClements} \affiliation{CCFE, Culham Science Centre, Abingdon, OX14 3DB, UK}
\author{S.C.~Chapman} \affiliation{Centre for Fusion, Space and Astrophysics, Warwick University, Coventry, CV4 7AL, UK}
\author{R.O.~Dendy} \affiliation{CCFE, Culham Science Centre, Abingdon, OX14 3DB, UK}\affiliation{Centre for Fusion, Space and Astrophysics, 
Warwick University, Coventry, CV4 7AL, UK}
\author{W.N.~Lai} \affiliation{Centre for Fusion, Space and Astrophysics, Warwick University, Coventry, CV4 7AL, UK}
\author{S.J.P. ~Pamela} \affiliation{CCFE, Culham Science Centre, Abingdon, OX14 3DB, UK}
\author{V.F.~Shevchenko} \affiliation{CCFE, Culham Science Centre, Abingdon, OX14 3DB, UK}
\author{R.G.L.~Vann} \affiliation{York Plasma Institute, University of York, York, YO10 5DQ , UK}
\vskip 0.25cm 

\date{\today}

\begin{abstract}
Recent measurements of microwave and X-ray emission during edge localised mode (ELM) activity in tokamak plasmas provide a fresh perspective on 
ELM physics. It is evident that electron kinetics, which are not incorporated in standard (fluid) models for the instability that drives ELMs, play a key role in the new observations. These effects should be included in future models for ELMs and the ELM cycle. The observed radiative effects paradoxically imply acceleration of electrons parallel to the magnetic field combined with rapid acquisition of perpendicular momentum. It is shown that this paradox can be resolved by the action of the anomalous Doppler instability which enables fast collective radiative relaxation, in the perpendicular direction, of electrons accelerated in the parallel direction by inductive electric fields generated by the initial ELM instability.  
    
\end{abstract}


\maketitle

The edge localised mode (ELM) \cite{ASDEX,Loarte,Kamiya} is a recurring plasma phenomenon driven by steep pressure gradients near the boundary of tokamak 
plasmas in high confinement (H-mode) regimes. A large ELM can deposit $\sim$10\% of the plasma stored energy onto material components, thereby causing 
significant erosion. This would be unacceptable in the future burning plasma experiment ITER \cite{Federici}; it is therefore vital to understand 
all aspects of the plasma physics of ELMs in order to predict and mitigate their effects in future experiments. The instability that initiates ELMs
is relatively well understood at the fluid level of description (see e.g. \cite{Huysmans}). However the observations reported here of 
electromagnetic radiation from plasmas with ELM activity indicate that the physics of this process incorporates electron kinetic effects that are not
included in fluid descriptions, and occur on relatively short lengthscales and timescales. For example, in the Mega Amp Spherical Tokamak (MAST) 
\cite{MAST}, ELMs are accompanied by intense bursts of microwave emission (BMEs) in the electron cyclotron (EC) frequency range. These can reach 
brightness temperatures 3-4 orders of magnitude higher than the thermal background and detailed experimental accounts of this emission have been given by others \cite{Fuchs,Taylor,Austin,Bartlett} and new observations on MAST using the Synthetic Aperture Imaging Radiometer (SAMI) \cite{Shevchenko} are broadly consistent with these accounts. The peak emission frequency of the BMEs lies above the cyclotron frequency in the outer plasma edge, $\Omega_{\mathrm{c,\, edge}}$, with $\omega/\Omega_{\mathrm{c,\, edge}} \sim 1.3 - 1.4$ and with a bandwidth of $\delta \omega/\omega \sim 0.3$. This narrow bandwidth rules out an explanation of the BMEs in terms of highly relativistic electrons, unless some effect causes the emitting electrons to have a narrow range of energies \cite{ripple}. There is no external heating source on MAST capable of accelerating electrons in the perpendicular direction, so the inferred presence of energetic electrons must be due to an intrinsic plasma effect. 

In this Letter we show for the first time that BMEs in MAST can be attributed to the presence of suprathermal, magnetic field-aligned electron 
populations that drive waves in the EC range via the anomalous Doppler instability (ADI). An electron with velocity parallel to the magnetic field 
$v_{\parallel}$ is in resonance with a wave of frequency $\omega$ and parallel wavevector $k_{\parallel}$ if
\begin{equation}
\omega = \ell \Omega_{e} + k_{\parallel}v_{\parallel},  
\end{equation}
where $\ell$ is an integer and $\Omega_e$ is the EC frequency. The ADI occurs when waves are excited via this resonance condition with $\ell < 0$; 
in applications of the ADI to tokamak plasmas, the case $\ell = -1$ is generally found to be relevant \cite{Lai}. As the ADI proceeds, the 
suprathermal electrons acquire perpendicular momentum 
comparable to their parallel momentum on timescales of a few hundred EC periods, $\tau_c$. While predominantly electrostatic, the
waves can be converted to electromagnetic waves, and thus propagate freely to antennas outside the plasma \cite{Ram}. 

\begin{figure}[ht]
\includegraphics[clip=true, trim = 0.0cm 0.0cm 0.0cm 0.5cm, width = \columnwidth]{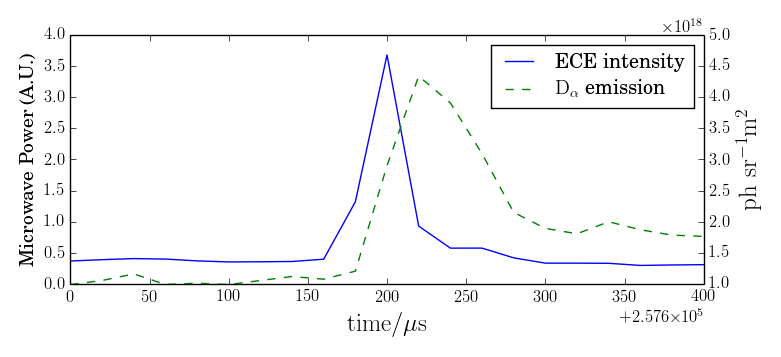}
\caption{\label{fig:example} Evolution of microwave intensity (solid) and D$_{\alpha}$ (dashed) emission in a typical ELM.}
\end{figure}

\begin{figure}[ht]
\includegraphics[clip=true, trim = 0.0cm 0.0cm 0.0cm 0.5cm, width = \columnwidth]{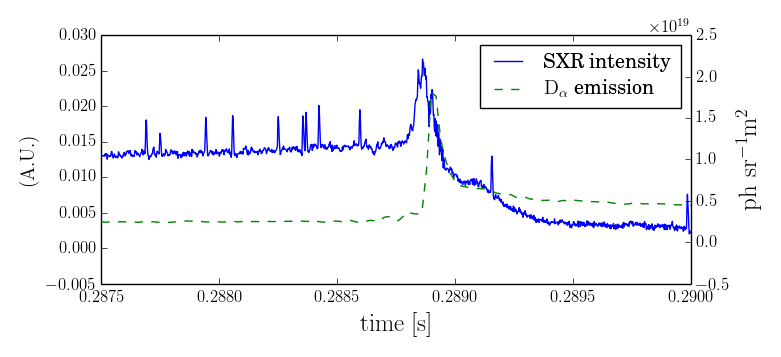}
\caption{\label{fig:SXR} Evolution of soft X-ray intensity (top) and D$_{\alpha}$ emission (bottom) in a typical ELM.}
\end{figure}

BME intensities relative to the average thermal background level peak at about 35dB and extend up to 40 dB. The BMEs also correlate in time with 
individual ELMs, as shown in Fig. \ref{fig:example}. The peak of the BME emission typically occurs 20$\mu$s before that of the deuterium-$\alpha$ 
(D$_{\alpha}$) line emission, which is the primary diagnostic signature of an ELM and is due to filamentary structures erupting into the relatively 
cold region outside the confined plasma. In many cases a burst of X-ray emission coincides with the microwave burst, with the peak emission again 
preceding the D$_{\alpha}$ emission by around 20$\mu$s in the crash phase \cite{kirk}. Figure \ref{fig:SXR} shows an example, obtained using a soft X-ray camera that is sensitive to photons with energies from 1keV to 30keV emitted along many lines-of-sight crossing the plasma. The fluxes plotted in Fig. \ref{fig:SXR} were recorded for a line-of-sight that passes through the plasma low field side edge above the midplane. The peak X-ray intensity is more than double the pre-ELM value. Smaller enhancements in X-ray intensity are seen for lines-of-sight passing closer to the plasma center.     

\begin{figure}[ht]
\includegraphics[clip=true, trim = 0.0cm 0.0cm 0.0cm 0.5cm, width = \columnwidth]{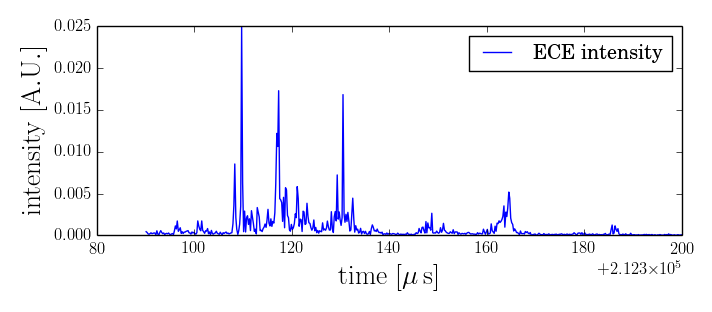}
\caption{\label{fig:zoomed_elm} Microwave intensity of a BME versus time recorded using high time resolution radiometer.}
\end{figure}

Radiometer measurements at high time resolution reveal fine structure in the BMEs, as shown in Fig. \ref{fig:zoomed_elm}. Individual bursts have
rise and decay times of the order of 1$\mu$s, and fine structure is also seen in the soft X-ray bursts (Fig. \ref{fig:SXR}).    

The near-universal occurrence of the BMEs, and the frequent concurrence of soft X-ray bursts, during ELMs in MAST, is strong evidence for the presence of energetic electrons. A rise in soft X-ray emission could arise from an inflow of impurities into the plasma as a result of the ELM. However, the fact that the peak in the soft X-ray emission precedes that of the D$_{\alpha}$ emission, with the X-ray emission decaying on a timescale of a few tens of microseconds, makes it more likely that the enhanced emission results from non-thermal bremsstrahlung.

\begin{figure}[ht]
\includegraphics[clip=true, trim = 5cm 5cm 2.5cm 5cm, width = 0.7\columnwidth]{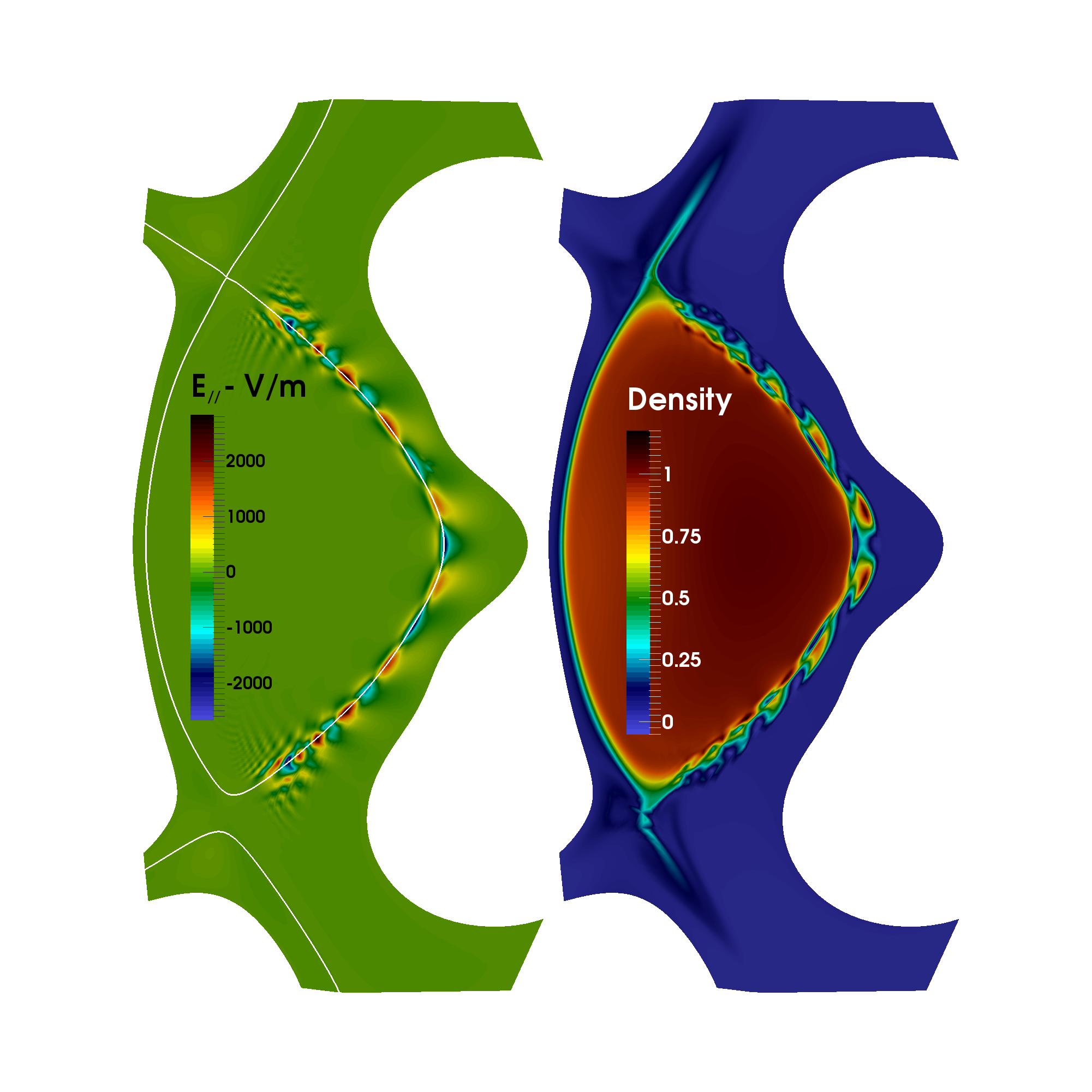}
\caption{\label{fig:EPar} Poloidal snapshots of parallel electric field (left plot) and plasma density normalised to the magnetic axis value (right 
plot) in a JOREK simulation of an ELM in MAST \cite{jorek}. The white curve in the left plot shows the separatrix.}
\end{figure}

While ELMs are likely to be triggered by ideal magnetohydrodynamic (MHD) instabilities, the later stages of these events, culminating in the
detachment of filamentary structures from the main plasma, must involve non-ideal effects. Simulations of ELMs in MAST performed with the nonlinear 
resistive MHD code JOREK \cite{jorek} reveal that parallel electric fields of up to 2kVm$^{-1}$ can be expected in localised regions (Fig. 
\ref{fig:EPar}). Mirnov coil measurements during ELMs in MAST imply that the modes associated with these fields have periods of the order of a few 
microseconds; individual electrons would be accelerated to tens of keV if they encountered 2kVm$^{-1}$ parallel electric fields for only a small 
fraction of this time. 

Given the presence of magnetic field-aligned suprathermal electrons, we now demonstrate that they are unstable to the generation of waves in the EC range under MAST like plasma conditions. To address this, particle-in-cell (PIC) simulations have been carried out in one space dimension and three velocity dimensions, the magnetic field being tilted at an angle of 45$^{\circ}$ with respect to the space axis. Thus, any excited waves have wavevectors with components parallel ($k_{\parallel}$) and perpendicular ($k_{\perp}$) to {\bf B} of equal magnitude, i.e. $k_{\parallel} = k_{\perp}$. Across much of the MAST plasma, due to the high $\beta$, the EC frequency is smaller than the electron plasma frequency $\omega_{pe}$; in conventional tokamaks the opposite inequality generally applies \cite{Lai}. The location in the MAST plasma of the instability producing the microwave bursts is not known precisely. The close correlation with ELMs 
suggests that it is likely to be close to the low field side plasma edge, but the electron density typically varies by more than an order of magnitude between the inner edge of the H-mode pedestal (steep pressure gradient region) and the plasma edge. For this reason, PIC simulations have been carried out with a range of values of initial bulk electron density $n_e$, from $2\times 10^{18}$m$^{-3}$ to $4\times 10^{19}$m$^{-3}$. In all cases the initial electron distribution comprises a Maxwellian bulk with a temperature $T_e$ of either 100eV or 20eV and a flat magnetic field-aligned tail, containing 5\% or 10\% of the total electron population and extending to 20 or 40 times $v_B$ where $v_B = (2T_e/m_e)^{1/2}$ is the initial bulk electron thermal speed, $m_e$ being the electron mass. It is known that the growth-rate of the ADI is linearly dependant on the fractional density $n_f/n_e$, so we may extrapolate the measured growth rates to smaller fractional densities if necessary. In all cases the equilibrium magnetic field $B_0$ is 0.4T, a typical value in the 
outer midplane edge of MAST plasmas.              

\begin{figure}[ht]
\includegraphics[clip=true, trim = 0.0cm 0.0cm 2.8cm 0.0cm, width = 0.8\columnwidth]{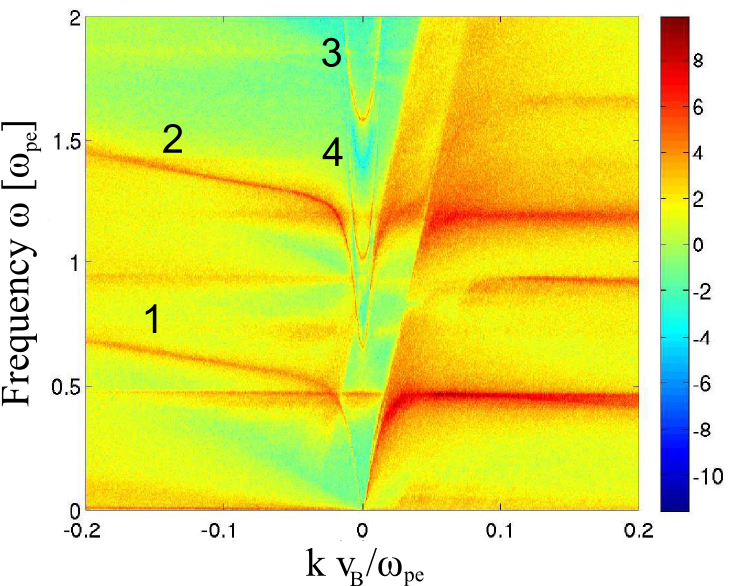}
\caption{\label{fig:dispersion_relation} Electrostatic field amplitude (on logarithmic scale) versus wavenumber and frequency in PIC simulation
with initial electron density $2\times 10^{18}$m$^{-3}$. Branch 1 is a whistler mode, 2 is a generalised langmuir mode, 3 is an O-mode, and 4 is an 
X-mode}
\end{figure}

Fig. \ref{fig:dispersion_relation} shows the result of Fourier transforming, in space and time, the electric field component parallel to the wavevector (i.e. the electrostatic component) in a simulation with initial $n_e$ equal to $2\times 10^{18}$m$^{-3}$. Two dominant forward-propagating ($k > 0$) waves are excited by the ADI, with frequencies at large $k$ approximately equal to 0.46$\omega_{pe} \simeq 2\pi \times 5.8$GHz and 1.2$\omega_{pe} \simeq 2\pi \times 15.2$GHz. The first of these frequencies is below the range detectable using microwave diagnostics on MAST while the latter agrees well with the observed peak emission frequencies for BMEs on MAST, and is also close to the upper hybrid resonance (UHR) frequency $\omega_{\mathrm{UH}} = (\omega_{pe}^2+\Omega_e^2)^{1/2}$ which is around 1.33$\omega_{pe}$ for the parameters used here. Predominantly electrostatic waves in this frequency range can be converted to electromagnetic modes, and then detected by antennas outside the plasma.

\begin{figure}[ht]
\includegraphics[clip=true, trim = 2.0cm 9.0cm 0.0cm 2.0cm, width = 10.0cm]{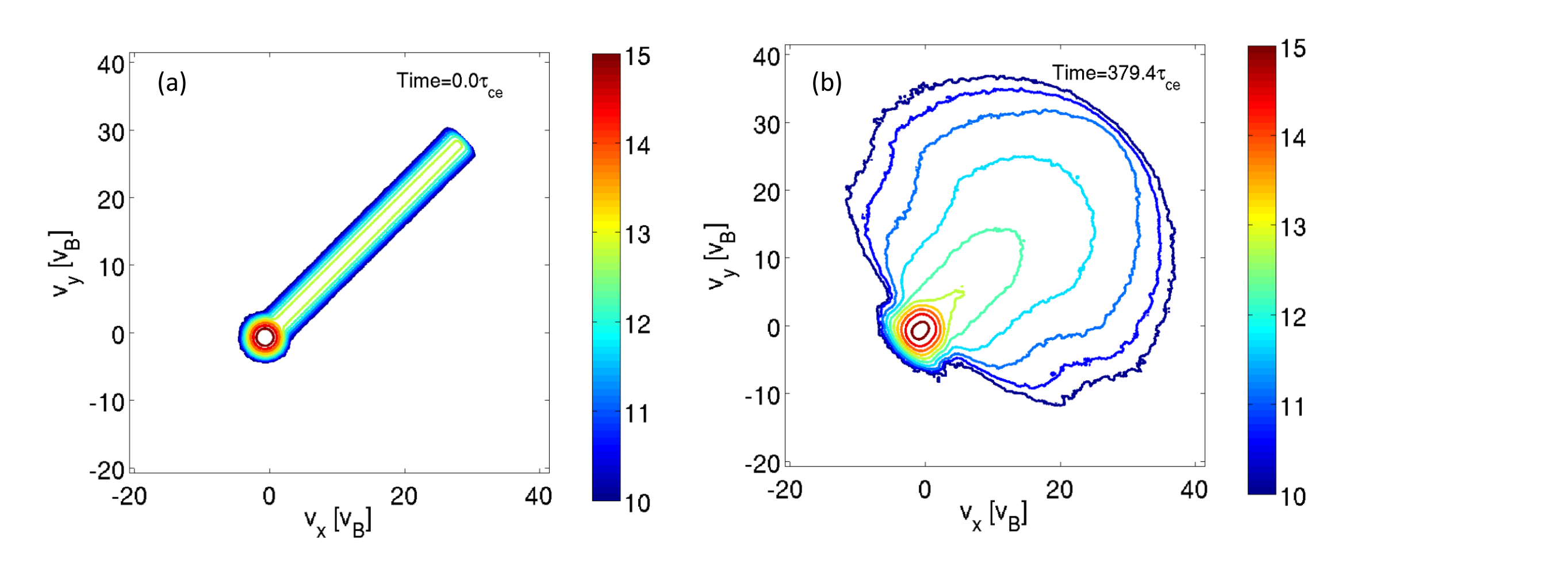}
\caption{\label{fig:electron_distribution} Contours of (a) initial and (b) final electron velocity distribution in PIC simulation with initial 
electron density $2\times 10^{18}$m$^{-3}$ and nonthermal electron tail fraction of 5\%. The colour bar has a logarithmic scale.} 
\end{figure}

Fig. \ref{fig:electron_distribution} shows (a) the initial electron velocity distribution in this simulation and (b) the distribution after 
379.4$\tau_c \simeq 30$ns. It is apparent that, in this short time, the electrons in the tail acquire perpendicular momenta comparable to their
initial parallel momenta. We conclude that strongly field-aligned energetic electron distributions in the edge plasma of MAST are very rapidly 
isotropised via the ADI. From Fig. \ref{fig:dispersion_relation} we note that the forward-propagating waves in the upper hybrid frequency range have
wavevectors $k \sim 0.1\omega_{pe}/v_B$ and hence $k_{\parallel} \sim 0.1\omega_{pe}/(\sqrt{2}v_B)$. Using this value of $k_{\parallel}$ in Eq. (1) and setting $\omega = 1.2\omega_{pe}$, $\ell = -1$ we find that the anomalous Doppler resonance condition is satisfied by electrons with $v_{\parallel} \sim 30v_B$; we note from Fig. \ref{fig:electron_distribution} that this is close to the parallel velocity at which electrons have acquired the largest boost in $v_{\perp}$, and conclude from this that the ADI explains the excitation of the high amplitude waves observed in this simulation.        

Similar phenomena are observed in PIC simulations with higher $n_e$ ($\ge 10^{19}$m$^{-3}$) and $T_e$ (100eV), more characteristic of the top of the
H-mode pedestal region in MAST plasmas. However, as in the low density case discussed above, high amplitude waves are only excited in these simulations at frequencies either below $\Omega_e$ or above $\omega_{pe}$. Neither of these frequency ranges is compatible with the observed frequency peak, if $B_0 = 0.4$ and $n_e \ge 10^{19}$m$^{-3}$. Only in the low density simulation do we see wave excitation at frequencies that are consistent with the BME spectra, suggesting that the energetic electrons causing the emission undergo acceleration very close to the MAST plasma edge. This conclusion is bolstered by the predictions from JOREK shown in Fig. \ref{fig:EPar}.

A corollary of this last conclusion is that the energetic electrons are likely to have a very short confinement time, particularly in view of the
fact that the magnetic field, \textbf{B}, in the plasma edge region is strongly distorted by the ELM instability. In these circumstances we expect the electrons to be subject to rapid cross-field transport of the type discussed by Rechester and Rosenbluth \cite{RR}. Generally, the condition 
$\nabla\cdot{\bf B} = 0$ implies that a perturbation $\delta{\bf B}$ to a magnetic field {\bf B} causes the field line to be displaced by $\Delta \sim (\delta B/B)L_c$ where $L_c$, the parallel length scale of the field, can be taken in tokamaks to be of order $qR$ where $q$ is the safety factor
and $R$ is major radius. This implies a cross-field diffusion rate $D_e$ of order $\Delta^2/\tau_{\parallel}$ where $\tau_{\parallel} \sim qR/v_{\parallel}$, i.e. $D_e \sim v_{\parallel}qR(\delta B/B)^2$, and a confinement time $\tau_e \sim \delta r^2/D_e = \delta r^2(B/\delta B)^2/(qRv_{\parallel})$ where $\delta r$ is the radial distance from the acceleration site to the last closed flux surface. Given that $v_{\parallel}$ is likely to be of order $10^8$ms$^{-1}$, while $R \simeq 1\,$m, $q \gg 1$ and $\delta r \sim 1\,$cm, it is evident that confinement times of the order of microseconds, consistent with the rapid burst timescales apparent in Fig. \ref{fig:zoomed_elm}, could result from rather modest field perturbations, e.g. $\delta B/B \sim 10^{-4}$.    
          
We consider finally the soft X-ray data. It is reasonable to suppose that the pre-burst emission is dominated by thermal bremsstrahlung, while the 
emission during the burst consists largely of non-thermal bremsstrahlung produced by energetic electrons. Using the simplest approximation to the 
bremsstrahlung cross section \cite{Kramers}, assuming that the energetic electrons have a flat tail distribution of the type used in the PIC simulations, and integrating the bremsstrahlung energy flux over photon energies above the soft X-ray camera threshold ${\cal E}_0$, the ratio of non-thermal to thermal emission from a homogeneous region of plasma is given by      
\begin{equation}
{{\cal F}_{\rm NT}\over {\cal F}_{\rm T}} = {\pi^{1/2}\over 3}{n_f\over n_e}\left({E_{\rm max}\over T_e}\right)^{1/2}e^{{\cal E}_0/T_e},  
\end{equation}
where $n_f$ and $E_{\rm max}$ are the density and maximum energy of the energetic electrons. As discussed previously, Fig. \ref{fig:SXR} shows fluxes of X-rays emitted along a chord passing through the plasma edge region, where $T_e$ is at most 100-200 eV. Since ${\cal E}_0 \simeq 1\,$keV, the exponential factor in Eq. (2) is much larger than unity and, if we assume also that $E_{\rm max}$ lies in the tens of keV range, it follows that a relatively modest energetic electron fraction $n_f/n_e$ (typically 10$^{-3}$ or less) would be sufficient to produce the enhancement in X-ray intensity apparent in Fig. \ref{fig:SXR}. The inference that only a small fraction of the local electron population is accelerated during an ELM is reinforced by Thomson scattering (TS) measurements of $T_e$. Large transient rises in the edge plasma $T_e$ have not been detected when TS measurements coincide with ELMs.                       

In conclusion, measurements of microwave and soft X-ray emission during ELMs in spherical tokamak plasmas provide strong evidence for the transient presence in the edge plasma of highly suprathermal electrons. The paradoxical nature of the intensity of the emission, given the absence of any RF wave heating is resolved by considering the response of electrons accelerated by a parallel electric field and invoking the collective anomalous Doppler instability. Particle-in-cell simulations show that magnetic field-aligned energetic electron distributions, of the kind inferred to result from parallel electric fields generated by ELMs, excite electromagnetic waves in the electron cyclotron range. Further, the PIC simulations predict emission frequencies between ${\omega_\mathrm{pe}}$ and $\omega_{\mathrm{UH}}$, this is in turn consistent with the predictions from JOREK about the radial location of the accelerating field. While soft X-ray and Thomson scattering data indicate that the fraction of accelerated electrons is small, their active role suggests that purely fluid models of ELMs are incomplete. For example the radial current associated with the rapid radial transport of these electrons could have an effect on ELM dynamics, and, if sufficiently energetic and present in sufficiently large numbers, they could cause damage to plasma-facing components. For these reasons it is hoped that the present study will prompt further investigations of energetic electron production during ELMs.

%
This project has received funding from the European Union's Horizon 2020 research and innovation programme under 
grant agreement number 633053 and from the RCUK Energy Programme [grant number EP/I501045]. Assistance from Luca 
Garzotti and Rory Scannell on the interpretation of soft X-ray and Thomson scattering data is 
gratefully acknowledged. To obtain further information on the data and models underlying this paper please contact 
PublicationsManager@ccfe.ac.uk. The views and opinions expressed herein do not necessarily reflect those of the 
European Commission.
%

\end{document}